\documentclass[fleqn,usenatbib]{mnras}
\usepackage{newtxtext,newtxmath}
\usepackage[T1]{fontenc}
\usepackage{ae,aecompl}

\usepackage{graphicx}	
\usepackage{amsmath}	
\usepackage{subcaption}
\usepackage{bm}
\usepackage[utf8]{inputenc}

\newcommand{\subB}{_{_{\rm B}}}
\newcommand{\subBFIT}{_{_{\rm B:FIT}}}
\newcommand{\subCON}{_{_{\rm CONFIG}}}
\newcommand{\subD}{_{_{\rm D}}}
\newcommand{\subF}{_{_{\rm F}}}

\newcommand{\subFIT}{_{_{\rm FIT}}}
\newcommand{\subHB}{_{_{\rm H_2:B}}}
\newcommand{\subHO}{_{_{\rm H_2:O}}}
\newcommand{\subINT}{_{_{\rm INTRINSIC}}}

\newcommand{\subO}{_{_{\rm O}}}
\newcommand{\subOFIT}{_{_{\rm O:FIT}}}
\newcommand{\subTOT}{_{_{\rm TOT}}}

\begin{document} 

\title[The surface-density profiles of filaments]{A systematic bias in fitting the surface-density profiles of interstellar filaments}
\author[A. P. Whitworth, F. D. Priestley \& D. Arzoumanian]{A. P. Whitworth$^{1}$\thanks{E-mail: ant@astro.cf.ac.uk},
F. D. Priestley$^{1}$ and D. Arzoumanian$^{2}$\\
$^{1}$School of Physics and Astronomy, Cardiff University, Cardiff CF24 3AA, UK\\
$^{2}$Aix Marseille Universit{\'e}, CNRS, CNES, LAM, Marseille, France}
\date{Accepted XXX. Received YYY; in original form ZZZ}
\pubyear{2019}
\label{firstpage}
\pagerange{\pageref{firstpage}--\pageref{lastpage}}
\maketitle

\begin{abstract}
The surface-density profiles of dense filaments, in particular those traced by dust emission, appear to be well fit with Plummer profiles, i.e. $\Sigma(b)=\Sigma\subB+\Sigma\subO\{1+[b/w\subO]^2\}^{[1-p]/2}$. Here $\,\Sigma\subB$ is the background surface-density; $\;\Sigma\subB+\Sigma\subO$ is the surface-density on the filament spine; $\;b$ is the impact parameter of the line-of-sight relative to the filament spine; $\;w\subO$ is the Plummer scale-length (which for fixed $p$ is exactly proportional to the full-width at half-maximum, $w\subO=\mbox{\sc fwhm}/2\{2^{2/[p-1]}-1\}^{1/2}$); and $\,p$ is the Plummer exponent (which reflects the slope of the surface-density profile away from the spine). In order to improve signal-to-noise it is standard practice to average the observed surface-densities along a section of the filament, or even along its whole length, before fitting the profile. We show that, if filaments do indeed have intrinsic Plummer profiles with exponent $p\subINT$, but there is a range of $w\subO$ values along the length of the filament (and secondarily a range of $\Sigma\subB$ values), the value of the Plummer exponent, $p\subFIT$, estimated by fitting the averaged profile, may be significantly less than $p\subINT$. The decrease, $\Delta p\!=\!p\subINT\!-\!p\subFIT$, increases monotonically (i) with increasing $p\subINT$; (ii) with increasing range of $w\subO$ values; and (iii) if (but only if) there is a finite range of $w\subO$ values, with increasing range of $\Sigma\subB$ values. For typical filament parameters the decrease is insignificant if $p\subINT =2$ ($0.05\lesssim\Delta p\lesssim 0.10$), but for $p\subINT =3$ it is larger ($0.18\lesssim\Delta p\lesssim 0.50$), and for $p\subINT =4$ it is substantial ($0.50\lesssim\Delta p\lesssim 1.15$). On its own this effect is probably insufficient to support a value of $p\subINT$ much greater than $p\subFIT \simeq 2$, but it could be important in combination with other effects.
\end{abstract}


\section{Introduction}

In the last decade it has become clear that filaments play a critical role in assembling the material to form stars \citep[e.g.][]{SchneiderS+ElmegreenB1979, BallyJetal1987, AbergelAetal1994, CambresyL1999, MyersP2009, HacarA+TafallaM2011, PerettoNetal2012, HacarAetal2013, PalmeirimPetal2013, AndrePetal2014, AlvesDeOliveiraCetal2014, PerettoNetal2013, KonyvesVetal2015, MarshKetal2016, HacarAetal2017, WardThompsonDetal2017, WilliamsGetal2018, HacarAetal2018, WatkinsEetal2019, LadjelateBetal2020, ArzoumanianDetal2021}. Even in clouds that are not apparently forming stars, or forming them very slowly \citep[e.g.][]{JoncasG1992, FalgaroneEetal2001, McClureGriffithsNetal2006, WardThompsonDetal2010}, including The Brick in the Central Molecular Zone of the Galaxy \citep{FederrathCetal2016}, the internal structure is still dominated by filaments

Filaments are particularly pronounced in maps of thermal dust-emission, such as those made using the Herschel Space Telescope \citep[e.g.][]{AndrePetal2010, Men'shchikovAetal2010, MolinariSetal2010, HillTetal2011, HennemannMetal2012, SchneiderNetal2012, SchisanoEetal2014, BenedettiniMetal2015, WangKetal2015, CoxNetal2016}. Given maps of thermal dust emission at a range of different wavelengths, it is possible to derive maps of the dust optical-depth, $\tau\subF$ at a fiducial far-infrared wavelength, $\lambda\subF$, either by Modified Blackbody fitting \citep[e.g.][]{HillTetal2011, PerettoNetal2012, SchneiderNetal2012, PalmeirimPetal2013, AlvesDeOliveiraCetal2014, BenedettiniMetal2015, WangKetal2015, CoxNetal2016, LadjelateBetal2020}, or by more sophisticated techniques \citep[e.g.][]{HowardAetal2019, HowardAetal2021} like {\sc ppmap} \citep{MarshKetal2015, WhitworthAetal2019}. Such maps are more accurate if the range of wavelengths (a) is large, and (b) extends well above and well below the peak of the spectral energy distribution. It is also necessary that the emission at all the wavelengths used be optically thin.

If the mass opacity, $\kappa\subF$, of dust at $\lambda\subF$ is known (and universal), one can convert a map of $\tau\subF$ into a map of the surface-density of dust, $\Sigma\subD=\tau\subF/\kappa\subF$. If the fraction of dust by mass, $Z\subD$ is known (and universal), one can convert this map into a map of the total surface-density (hereafter simply `the surface-density), $\Sigma =\Sigma\subD/Z\subD$. Finally, if one assumes that all the hydrogen is molecular, one can convert the map of $\Sigma$ into a map of the column-density of molecular hydrogen, $N_{_{\rm H_2}}=X\Sigma/2m_{_{\rm H}}$. Here $X$ is the fraction of hydrogen by mass and $m_{_{\rm H}}$ is the mass of an hydrogen atom. With $X=0.70$ this reduces to
\begin{eqnarray}\label{EQN:NH2.2.Sigma.01}
N_{_{\rm H_2}}&=&4.4\times 10^{19}\,\rm{cm^{-2}}\left[\frac{\Sigma}{\rm{M_{_\odot}\,pc^{-2}}}\right].
\end{eqnarray}
However this last conversion neglects the fact that on most lines of sight a significant fraction of the hydrogen is not molecular. Therefore in the sequel we prefer to present our analysis in terms of $\Sigma$.

\subsection{Plummer profiles}

Dust-emission filaments are found to have surface-density profiles (hereafter SDPs) that can be fit with Plummer profiles,
\begin{eqnarray}\label{EQN:Sigma.b.1}
\Sigma(b)&=&\Sigma\subB\,+\,\Sigma\subO\left\{1+\left[\frac{b}{w\subO}\right]^2\right\}^{-[p-1]/2}
\end{eqnarray}
\citep[e.g.][]{ArzoumanianDetal2011, PalmeirimPetal2013, CoxNetal2016, AndrePetal2016, ArzoumanianDetal2019, HowardAetal2019, HowardAetal2021}. In Eq. \ref{EQN:Sigma.b.1}, $\Sigma\subB$ is the background surface-density, $\Sigma\subB +\Sigma\subO$ is the surface density on the filament spine, $b$ is the impact parameter of the line of sight relative to the filament spine (i.e. projected distance from the filament spine), $w\subO$ is the Plummer scale-length relating to the width of the densest part of the filament, and $p$ is the Plummer exponent relating to the density gradient in the outer parts of the filament.

The full-widths at half-maximum surface-density is related to $w\subO$ by
\begin{eqnarray}\label{EQN:fwhm.wO.1}
\mbox{\sc fwhm}&\simeq&2w\subO\left\{2^{2/[p-1]}\,-\,1\right\}^{1/2},
\end{eqnarray}
so for fixed $p$ the {\sc fwhm} is exactly proportional to $w\subO$. Implicitly the Plummer exponent is
\begin{eqnarray}
p&=&1\,-\,\mbox{\rm LIM}_{b\rightarrow\infty}\!\left\{\frac{d\ln(\Sigma -\Sigma\subB)}{d\ln(b)}\right\}.
\end{eqnarray}

Provided that (a) the contribution from the background ($\Sigma\subB$) is uniform,  and (b) the filament is cylindrically symmetric, the filament's underlying volume-density profile (VDP) should also subscribe to a Plummer profile, viz.
\begin{eqnarray}
\rho(w)&=&\rho\subB\,+\,\rho\subO\left\{1+\left[\frac{w}{w\subO}\right]^2\right\}^{-p/2},
\end{eqnarray}
as shown by \citet{CasaliM1986}. Here $\rho\subB$ is the background volume-density, $\rho\subO$ is the excess volume-density on the filament spine, $w$ is the true (i.e. 3D) radial distance from the filament spine, and $w\subO$ is the same Plummer scale-length as invoked in Eqs. \ref{EQN:Sigma.b.1} and \ref{EQN:fwhm.wO.1}. Implicitly the Plummer exponent is
\begin{eqnarray}
p&=&-\,\mbox{\rm LIM}_{w\rightarrow\infty}\!\left\{\frac{d\ln(\rho-\rho\subB)}{d\ln(w)}\right\},
\end{eqnarray}
and
\begin{eqnarray}
\rho\subO&=&\frac{\Sigma\subO\,\Gamma(p/2)\,\cos(\psi)}{w\subO\,\Gamma(1/2)\,\Gamma(p/2\!-\!1/2)};
\end{eqnarray}
$\Gamma$ is the gamma function and $\psi$ is the angle between the filament spine and the plane of the sky.

\subsection{Longitudinally averaged filament profiles}

When fitting SDPs with Plummer profiles (i.e. Eq. \ref{EQN:Sigma.b.1}), it is standard practice to first derive a single profile averaged along the length of the whole filament or a section thereof (in order to improve signal-to-noise), and then to estimate the Plummer parameters that best fit these longitudinally averaged filament profiles \citep[e.g.][]{ArzoumanianDetal2011, ArzoumanianDetal2019, HowardAetal2019, HowardAetal2021}. We distinguish parameters derived in this way with a subscript `{\sc fit}'.

$\Sigma_{_{\rm B:FIT}}$ and $\Sigma_{_{\rm O:FIT}}$ can be determined directly (modulo some straightforward interpolation). There are then only two further parameters to estimate: $w_{_{\rm O:FIT}}$ (or strictly speaking its angular equivalent, $\theta_{_{\rm O:FIT}}=w_{_{\rm O:FIT}}/D$, where $D$ is the distance to the source); and $p\subFIT$. Values of $w_{_{\rm O:FIT}}\sim 0.03\,{\rm pc}$ (corresponding to $\mbox{\sc fwhm} \sim 0.1\,{\rm pc}$) and $p\subFIT\sim 2$ are commonly reported for the filaments observed in local molecular clouds \citep[e.g.][]{ArzoumanianDetal2011, PalmeirimPetal2013, AndrePetal2014, ArzoumanianDetal2019}, and also for the filaments identified in hydrodynamic and magneto-hydrodynamic simulations of turbulent molecular clouds \citep[e.g.][]{SmithRetal2014, KirkHetal2015, FederrathC2016, PriestleyFWhitworthA2020}.

However, \citet{PanopoulouG2017} have pointed out that the distribution of intrinsic {\sc fwhm} values for interstellar filaments, although centred on $\,\sim\!0.1\,{\rm pc}$, may be significantly broader than reported, due to the averaging process.

Here we show that the intrinsic $p$ values of interstellar filaments (hereafter $p\subINT$) may also be larger than reported, i.e. $p\subINT > p\subFIT$, again due to the averaging process. Specifically, the reduction, $\Delta p = p\subINT - p\subFIT$ is larger for larger values of $p\subINT$. $\Delta p$ is also larger if the range of $w\subO$ values is larger. And finally, provided there is a finite range of $w\subO$ values, $\Delta p$ is larger if the range of $\Sigma\subB$ values is larger.

We define our computational methodology in \S\ref{SEC:CompMeth}. We present our results in \S\ref{SEC:Results}. We summarise our conclusions in \S\ref{SEC:Conc}.

\section{Computational Methodology}\label{SEC:CompMeth}

\subsection{Logarithmic box-car distributions}

In the sequel, SDPs are generated with the Plummer exponent held fixed at $p=p\subINT$. For each of the other three parameters defining the Plummer SDP (i.e. $X$ $\equiv$ $[\Sigma\subB/\rm{M_{_\odot}\,pc^{-2}}]$, $[\Sigma\subO/\rm{M_{_\odot}\,pc^{-2}}]$, $[w\subO/\rm{pc}]$; see Eqs. \ref{EQN:Sigma.b.1} and \ref{EQN:fwhm.wO.1}) we assume that $\log_{_{10}}\!(X)$ has a box-car distribution, with mean $\mu_{_X}$ and standard deviation $\sigma_{_X}$,\footnote{Strictly speaking, `$\mu_{_X}\!$' should read `$\mu_{_{\log_{10}(X)}}\!$', and `$\sigma_{_X}\!$' should read `$\sigma_{_{\log_{10}(X)}}\!$'. We use the shorter version for convenience.} i.e.
\begin{eqnarray}
\frac{dP}{d\!\log_{_{10}}\!(X)}\!&\!\!=\!\!&\!\left\{\begin{array}{ll}
\frac{d\!\log_{_{10}}\!(X)}{2\surd{3}\,\sigma_{_X}},\hspace{0.2cm}&|\log_{_{10}}\!(X)-\mu_{_X}|\leq\surd{3}\sigma_{_X};\\
0\,,&|\log_{_{10}}\!(X)-\mu_{_X}|>\surd{3}\sigma_{_X}.\hspace{0.4cm}\\
\end{array}\right.
\end{eqnarray}

\begin{table}
\begin{center}
\caption{The box-car distribution parameters (hereafter simply `the distribution parameters'). The lefthand column gives the fixed values of $\mu_{_X}$, which represents the mean of $\log_{_{10}}(X)$. The righthand column gives the range of $\sigma_{_X}$ that we explore, where $\sigma_{_X}$ represents the standard deviation of $\log_{_{10}}(X)$. $X$ stands for $[\Sigma\subB/\rm{M_{_\odot}\,pc^{-2}}]$, $[\Sigma\subO/\rm{M_{_\odot}\,pc^{-2}}]$ and $[w\subO/\rm{pc}]$.}
\begin{tabular}{ll}\hline
{\sc Fixed means}\hspace{2.0cm} &{\sc Ranges of standard deviations} \\\hline
\multicolumn{2}{l}{Background surface-density, $[\Sigma\subB/\rm{M_{_\odot}\,pc^{-2}}]$:} \\
$\mu_{_{\Sigma\subB}}=1.778\hspace{0.5cm}$ & $0\leq \sigma_{_{\Sigma\subB}}\leq 0.40$ \\\hline
\multicolumn{2}{l}{Excess surface-density on filament spine $[\Sigma\subO/\rm{M_{_\odot}\,pc^{-2}}]$.} \\
$\mu_{_{\Sigma\subO}}=1.778$ & $0\leq \sigma_{_{\Sigma\subO}}\leq 0.40$ \\\hline
\multicolumn{2}{l}{Plummer scale-length of filament, $[w\subO/\rm{pc}]$.} \\
$\mu_{_{w\subO}}=-1.523$ & $0\leq \sigma_{_{w\subO}}\leq 0.40$ \\\hline
\end{tabular}
\end{center}
\label{TAB:Params}
\end{table}

\subsection{Filament configurations}

For the purpose of this study, all the means ($\mu_{_{\Sigma\subB}}\!$, $\mu_{_{\Sigma\subO}}\!$, $\mu_{_{w\subO}}$) have fixed values, as given in the lefthand column of Table 1. These correspond to a median background surface-density, $\Sigma\subB=60\,\rm{M_{_\odot}\,pc^{-2}}$ (equivalently $N_{_{\rm H_2}}\simeq 2.6\times 10^{21}\,\rm{cm^{-2}}$), a median spinal surface-density $\Sigma\subO=60\,\rm{M_{_\odot}\,pc^{-2}}$, and a median Plummer scale-length $w\subO=0.03\,\rm{pc}$. These choices of $\mu_{_{\Sigma\subB}}$, $\mu_{_{\Sigma\subO}}$ and $\mu_{_{w\subO}}$ are informed by the results of \citet{ArzoumanianDetal2019}. 

The different filament configurations that we explore are therefore completely defined by specifying the three standard deviations: $\sigma_{_{\Sigma\subB}}\!$, $\sigma_{_{\Sigma\subO}}\!$ and $\sigma_{_{w\subO}}$. We consider values for these standard deviations within the limits specified in the righthand column of Table 1. A large standard deviation means that the associated parameter varies over a large range. The maximum standard deviations considered allow $\Sigma\subB$ and $\Sigma\subO$ to take values between $12$ and $300\,\rm{M_{_\odot}\,pc^{-2}}$; and $w\subO$ to take values between $0.006$ and $0.15\,\rm{pc}$. We are not suggesting that such extreme values of $\Sigma\subB$, $\Sigma\subO$ and $w\subO$ are the norm. Large values are treated in order to evaluate trends accurately. We discuss in Appendix \ref{APP:Arzoumanian} (see Table A1) the standard deviations that are actually observed.

\subsection{Multiple realisations and longitudinally averaged profiles}

For each filament configuration (i.e. each specific combination of $\sigma_{_{\Sigma\subB}}\!$, $\sigma_{_{\Sigma\subO}}\!$ and $\sigma_{_{w\subO}}$), we generate $c\subTOT\!=\!10^6$ different random combinations of $\Sigma\subB$, $\Sigma\subO$ and $w\subO$. For example, different values of $\Sigma\subO$ are obtained by generating linear random deviates, ${\cal L}$, on the interval $[0,1]$, and then setting
\begin{eqnarray}\label{EQN:SigmaO.L}
\Sigma\subO&=&10^{\mu_{_{\Sigma\subO}}\!+\,\surd{3}\sigma_{_{\Sigma\subO}}\![2{\cal L}-1]}.
\end{eqnarray}
Each combination of $\Sigma\subB$, $\Sigma\subO$ and $w\subO$ allows us to compute an individual SDP, and these individual SDPs are added and normalised to produce a longitudinally averaged SDP, $\Sigma\subFIT\!(b)$, for that filament configuration. The results presented below involve $\,\sim\!5\times 10^5$ different filament configurations, and hence $\,\sim\!5\times 10^{11}$ individual SDPs.

\subsection{Plummer profile fitting}

The average SDP, $\overline{\Sigma}\subCON\!(b)$, for a given filament configuration, $[\sigma_{_{\Sigma\subB}}\!, \sigma_{_{\Sigma\subO}}\!, \sigma_{_{w\subO}}\!]$, is fit with a Plummer-profile (Eq. \ref{EQN:Sigma.b.1}), and the best-fit parameters, ($\Sigma\subBFIT$, $\Sigma\subOFIT$ and $w\subOFIT$) are established to five significant figures. The quality of the fit is measured with the fractional root-mean-square error, ${\cal Q}\subFIT$, given by
\begin{eqnarray}\nonumber
{\cal Q}\subFIT^2\!\!&\!\!\!=\!\!\!&\!\!\frac{1}{i\subTOT}\sum\limits_{i=1}^{i=i\subTOT}\left\{\frac{1}{\overline{\Sigma}\subCON^2\!(b_i)}\left[\overline{\Sigma}\subCON\!(b_i)-\Sigma\subBFIT\vphantom{\left\{1\!+\!\left[\frac{b_i}{w\subOFIT}\right]^2\right\}^{-[p\subFIT-1]/2}}\right.\right.\\\label{EQN:Q1}
\!\!\!&\!\!\!\!&\!\!\!\hspace{1.5cm}\left.\left.-\,\Sigma\subOFIT\left\{1\!+\!\left[\frac{b_i}{w\subOFIT}\right]^2\right\}^{-[p\subFIT-1]/2}\right]^2\right\}\!.\hspace{0.5cm}
\end{eqnarray}
Here the $b_i$ $\;(i\!=\!1$ to $i\subTOT\!=\!401)$ are impact parameters uniformly spaced between $b_{_1}\!=\!0.000\,{\rm pc}$ and $b_{_{401}}\!=\!0.400\,{\rm pc}$.

\subsection{Correlated Plummer parameters}

\citet{ArzoumanianDetal2019} note that the surface-density on the filament spine is correlated with the background surface-density (see their Figure 6c and the associated caption). Specifically they find
\begin{eqnarray}
N\subHO\!\!&\!\simeq\!&\!\![0.95\pm 0.15]N\subHB-[0.15\pm 0.39]\!\times\!10^{21}\,\rm{cm^{-3}},\hspace{0.7cm}
\end{eqnarray}
where $N\subHB\!+\!N\subHO$ is the column-density of molecular hydrogen on the filament spine, and $N\subHB$ is the column-density of molecular hydrogen in the background. We have therefore repeated our analysis with the equation for generating values of $\Sigma\subO$ (i.e. Eq. \ref{EQN:SigmaO.L}) replaced by the equivalent equation in terms of surface density:
\begin{eqnarray}\label{EQN:correlation01}
\Sigma\subO&=&\left\{0.95\,\Sigma\subB\,-\,3.4\,\rm{M_{_\odot}\,pc^{-2}}\right\}\;10^{\surd{3}\sigma_{_{\Sigma\subO}}\![2{\cal L}-1]}.
\end{eqnarray}
In Appendix \ref{APP:Arzoumanian} we discuss the \citet{ArzoumanianDetal2019} data set in more detail, and possible reasons for this correlation. We do not consider correlations between any of the other pairs of Plummer distribution parameters.

\begin{figure}
\vspace{-1.20cm}\hspace{-0.90cm}\includegraphics[angle=0.0,width=1.27\columnwidth]{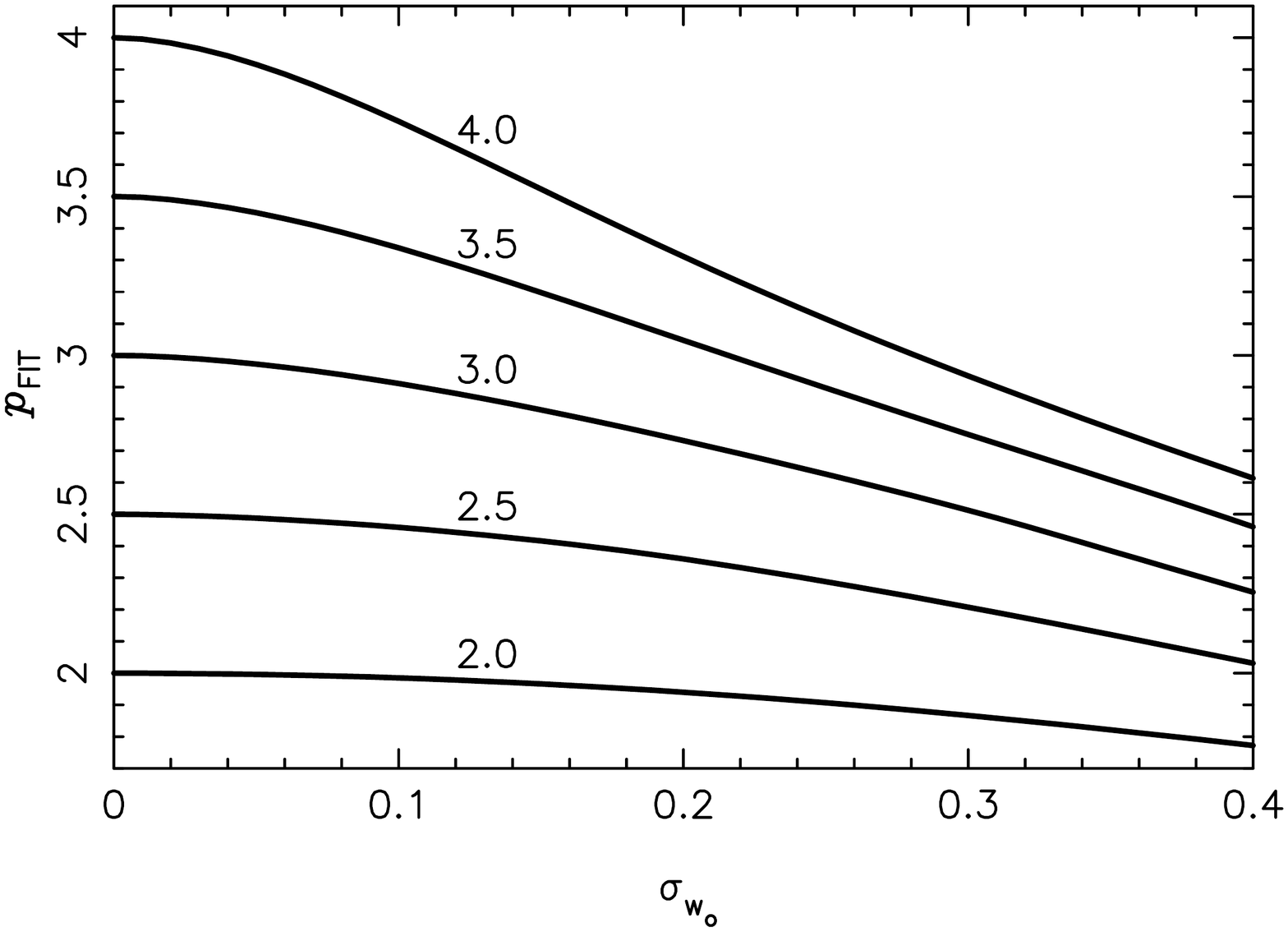}
\vspace{-0.5cm}
\caption{The variation of $p\subFIT$ with $\sigma_{_{w\subO}}$ when the other ranges are set to zero ($\sigma_{_{\Sigma\subB}}=\sigma_{_{\Sigma\subO}}=0$). Results are shown for $p\subINT =4.0,\;3.5,\;3.0,\;2.5\;{\rm and}\;2.0$, as labelled. For each curve 401 profiles have been generated and fitted. For the $p\subINT$ $=$ $4.0$, $3.5$, $3.0$, $2.5$ and $2.0$ curves the average fractional root-mean-square error is $\bar{\cal Q}$ $=$ $0.012(\pm 0.011)$, $0.012(\pm 0.010)$, $0.010(\pm 0.008)$, $0.008(\pm 0.005)$ and $0.006(\pm 0.004)$, respectively (see Eq. \ref{EQN:Q1}).}
\label{FIG:FilProf_pFIT_sgwO}
\end{figure}

\begin{figure}
\vspace{-1.20cm}\hspace{-0.9cm}\includegraphics[angle=270.0,width=1.27\columnwidth]{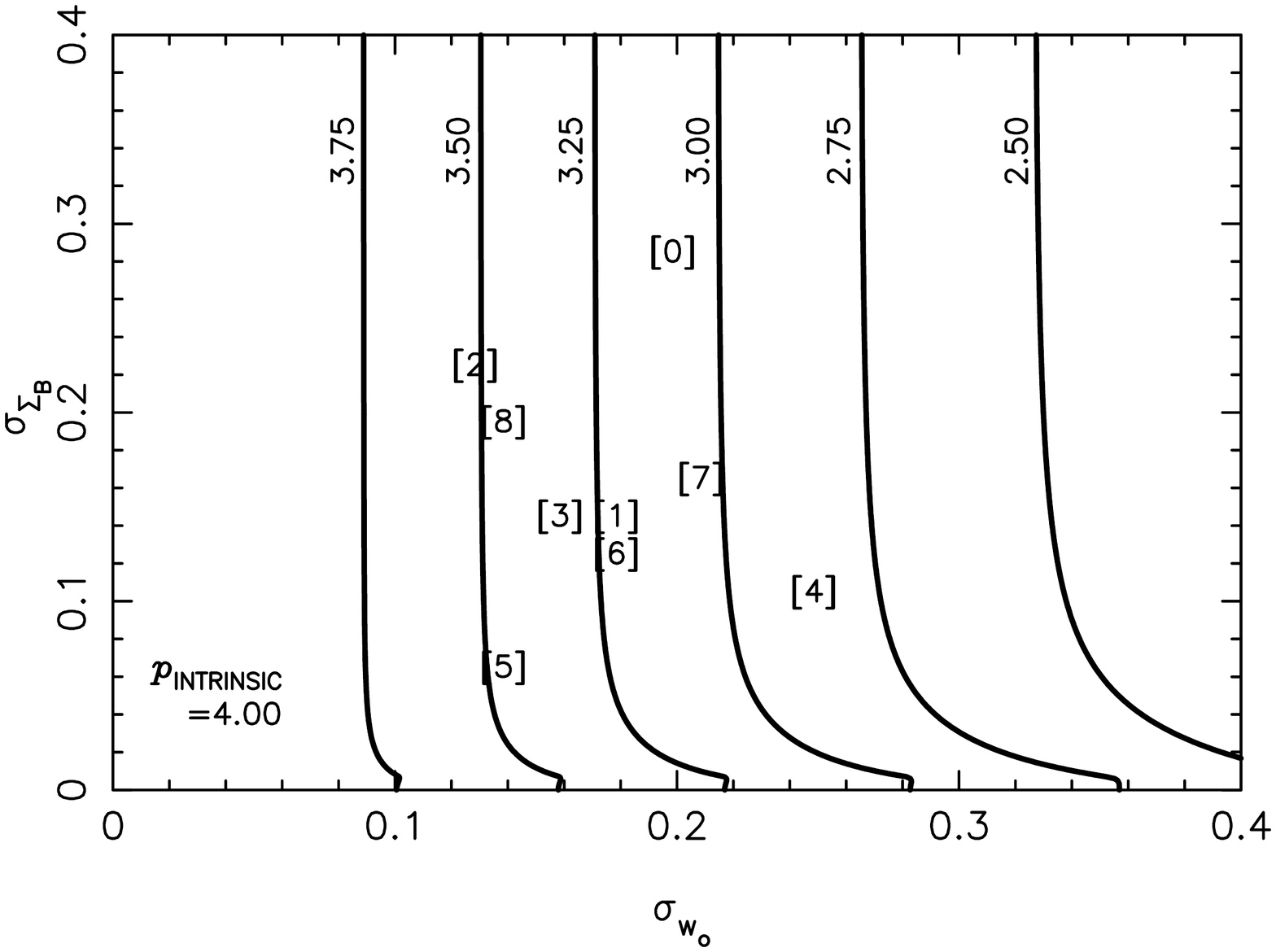}
\vspace{-0.5cm}
\caption{Contours of constant $p\subFIT$ on the $[\!\sigma_{_{w\subO}},\!\sigma_{_{\Sigma\subB}}\!\!]$ plane for $p\subINT = 4$. The $1.6\times 10^5$ profiles generated and fitted for this plot have an average fractional root-mean-square error $\bar{\cal Q}=0.0012(\pm 0.0034)$ (see Eq. \ref{EQN:Q1}). The numbers in square brackets represent the values for filaments in different regions: [0] All regions; [1] IC5146; [2] Orion B; [3] Aquila; [4] Musca; [5] Polaris; [6] Pipe; [7] Taurus L1495; [8] Ophiuchus (see Appendix A for details).}
\label{FIG:FilProf_pFIT_sgwO_sgSigB_pINT4}
\end{figure}

\begin{figure}
\vspace{-1.20cm}\hspace{-0.9cm}\includegraphics[angle=270.0,width=1.27\columnwidth]{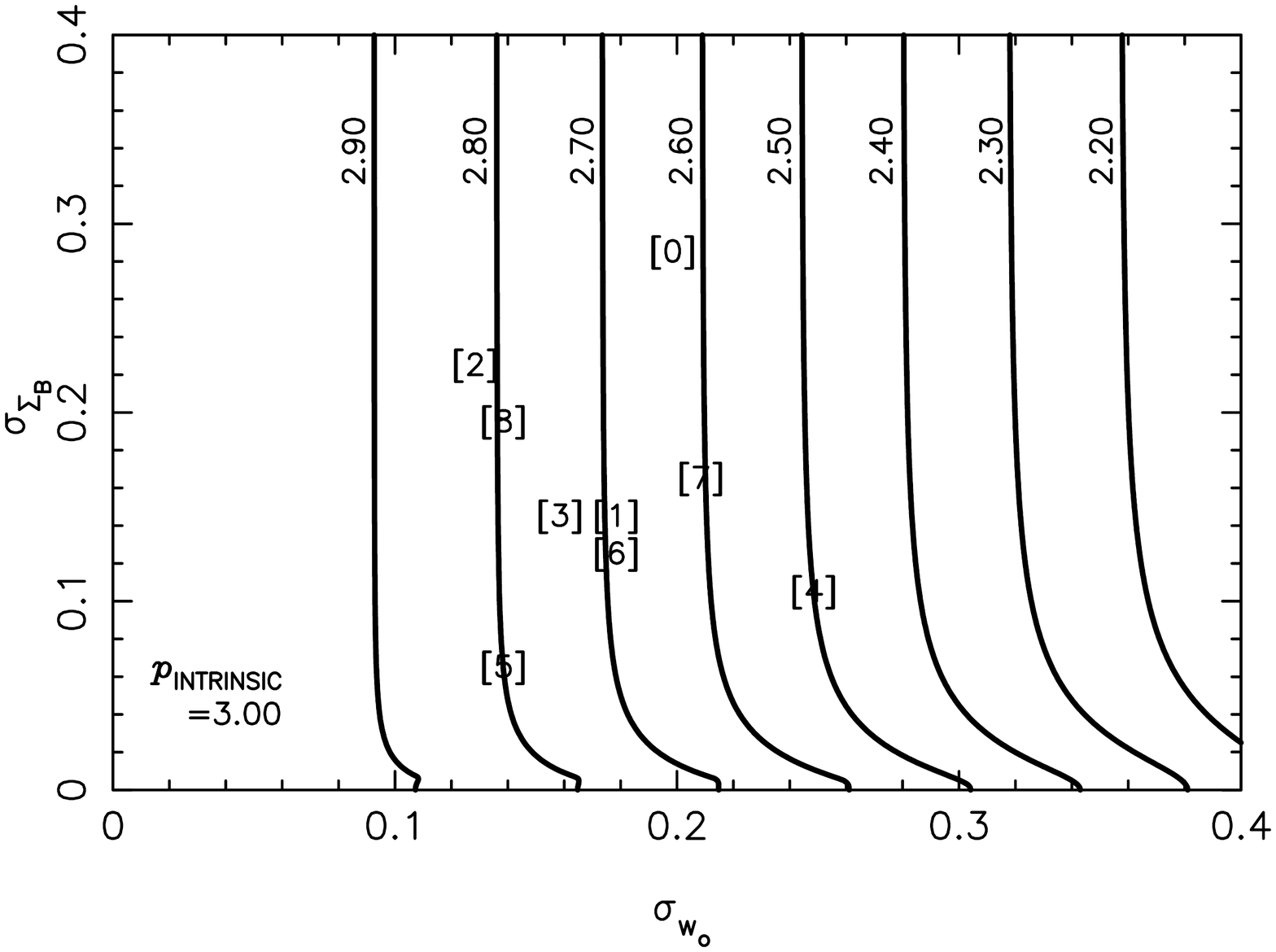}
\vspace{-0.5cm}
\caption{As Fig\,\ref{FIG:FilProf_pFIT_sgwO_sgSigB_pINT4}, but for $p\subINT =3$; $\bar{\cal Q}=0.0011(\pm 0.0028)$.}
\label{FIG:FilProf_pFIT_sgwO_sgSigB_pINT3}
\end{figure}

\begin{figure}
\vspace{-1.20cm}\hspace{-0.9cm}\includegraphics[angle=270.0,width=1.27\columnwidth]{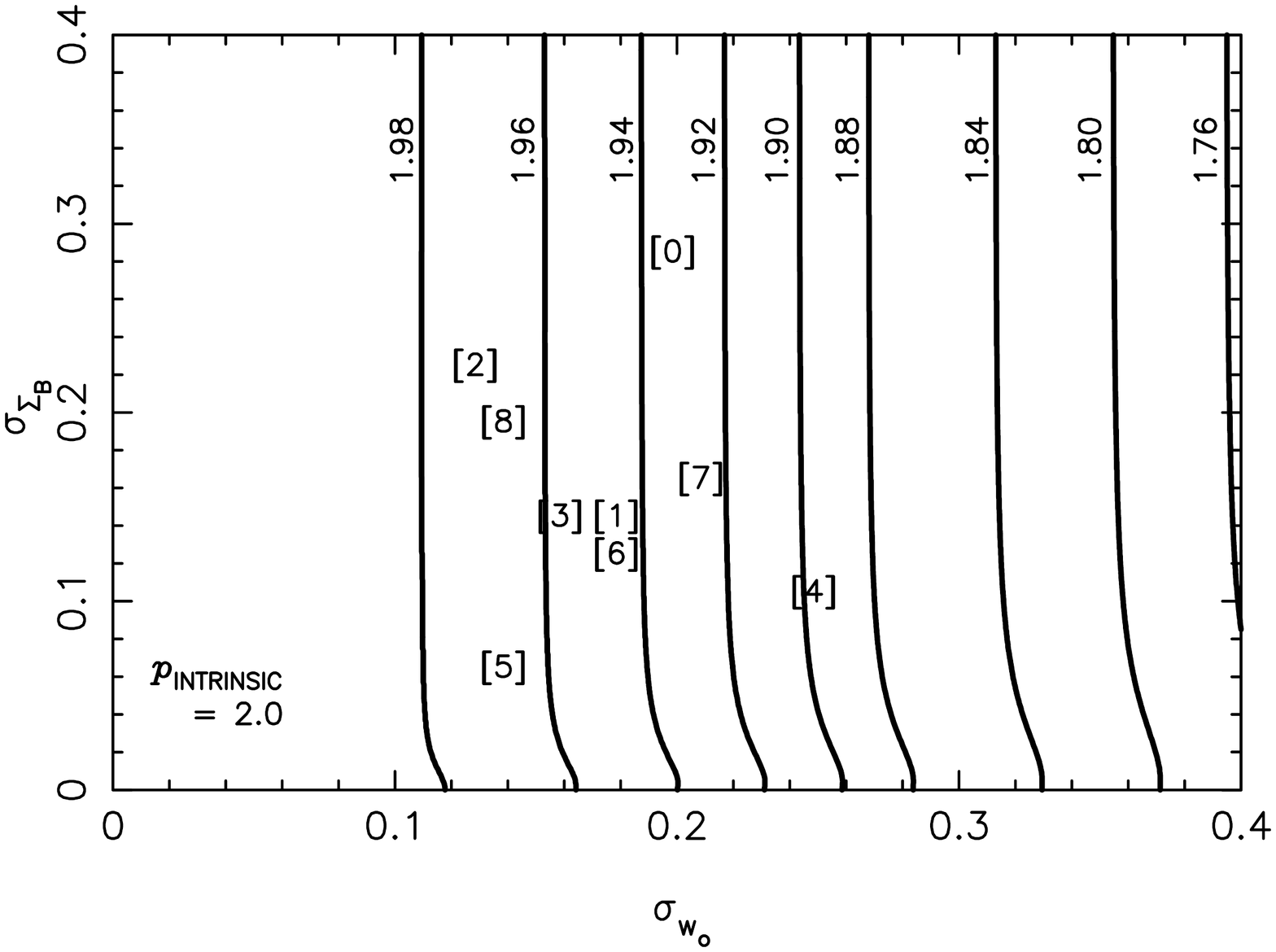}
\vspace{-0.5cm}
\caption{As Fig\,\ref{FIG:FilProf_pFIT_sgwO_sgSigB_pINT4}, but for $p\subINT =2$; $\bar{\cal Q}=0.0006(\pm 0.0016)$.}
\label{FIG:FilProf_pFIT_sgwO_sgSigB_pINT2}
\end{figure}

\section{Results}\label{SEC:Results}

We are concerned here with the values of $p\subFIT$ for a longitudinally averaged SDP when there are significant variations in the background surface-density, $\Sigma\subB$, and/or the spinal surface-density, $\Sigma\subO$, and/or the radial scale-length, $w\subO$, along the filament or section of filament being considered.

We label $w\subO$ the {\it primary} parameter, because a finite range of $w\subO$ values  always produces a reduction in $p\subFIT$, irrespective of whether there is variation in $\Sigma\subB$ or $\Sigma\subO$. The reduction increases with increasing range (i.e increasing $\sigma_{_{w\subO}}$). 

We label $\Sigma\subB$ the {\it secondary} parameter, because a finite range of $\Sigma\subB$ values only produces a reduction in $p\subFIT$ when there is also a finite range of $w\subO$ values. The associated reduction is relatively small, and increases with both the range of $w\subO$ values and the range of $\Sigma\subB$ values (i.e. increasing $\sigma_{_{w\subO}}\!$ and increasing $\sigma_{_{\Sigma\subB}}\!$). The reduction associated with the range of $\Sigma\subB$ values tends to saturate at large $\sigma_{_{\Sigma\subB}}\!$

We label $\Sigma\subO$ the {\it null} parameter, because whatever the range of $\Sigma\subO$ values it has no effect on $p\subFIT$.

\subsection{One parameter at a time}

To demonstrate these dependences, we first consider one parameter at a time, and increase the range of that parameter while keeping the other two parameters fixed. In other words we vary one of $\sigma_{_{\Sigma\subB}}$, $\sigma_{_{\Sigma\subO}}$ and $\sigma_{_{w\subO}}$ in turn, and set the other two to zero.

If we increase $\sigma_{_{\Sigma\subB}}$ (i.e. we increase the range of $\Sigma\subB$), with $\sigma_{_{\Sigma\subO}}\!=\sigma_{_{w\subO}}=0$ (i.e. fixed $\Sigma\subO=60\,\rm{M_{_\odot}\,pc^{-2}}$ and fixed $w\subO=0.03\,\rm{pc}$), this has no effect on $p\subFIT$, which remains exactly equal to $p\subINT$.

Likewise, if we increase $\sigma_{_{\Sigma\subO}}$ (i.e. we increase the range of $\Sigma\subO$), with $\sigma_{_{\Sigma\subB}}\!=\sigma_{_{w\subO}}=0$ (i.e. fixed $\Sigma\subB=60\,\rm{M_{_\odot}\,pc^{-2}}$ and fixed $w\subO=0.03\,\rm{pc}$), this too has no effect on $p\subFIT$, which remains exactly equal to $p\subINT$.

However, if we increase $\sigma_{_{w\subO}}$ (i.e. we increase the range of $w\subO$), with $\sigma_{_{\Sigma\subB}}\!=\sigma_{_{\Sigma\subO}}=0$ (i.e. fixed $\Sigma\subB=60\,\rm{M_{_\odot}\,pc^{-2}}$ and fixed $\Sigma\subO=60\,\rm{M_{_\odot}\,pc^{-2}}$), $p\subFIT$ is reduced, as shown on Fig. \ref{FIG:FilProf_pFIT_sgwO}. $\,w\subO$ is therefore the primary parameter on the grounds that it is the only parameter whose variation, on its own, affects $p\subFIT$. Moreover, it affects $p\subFIT$ strongly if $p\subINT$ is large.

\subsection{Two parameters at a time}

Next we consider two parameters at a time, and vary the ranges of those two parameters simultaneously while keeping the value of the third parameter fixed at its default value. In other words, we vary two of $\sigma_{_{\Sigma\subB}}$, $\sigma_{_{\Sigma\subO}}$ and $\sigma_{_{w\subO}}$, and set the remaining one to zero.

If we vary the range of $\Sigma\subB$ (i.e. vary $\sigma_{_{\Sigma\subB}}\!$) and simultaneously vary the range of $\Sigma\subO$ (i.e. vary $\sigma_{_{\Sigma\subO}}\!$), with $w\subO$ fixed at its default value (i.e. $w\subO =0.03\,\rm{pc}$), the value of $p\subFIT$ is unaffected, and remains at $p\subINT$.

If we vary the range of $\Sigma\subB$ (i.e. vary $\sigma_{_{\Sigma\subB}}\!$) and simultaneously vary the range of $w\subO$ (i.e. vary $\sigma_{_{w\subO}}$), with $\Sigma\subO$ fixed at its default value (i.e. $\Sigma\subO =60\,\rm{M_{_\odot}\,pc^{-2}}$), both variations produce a change in $p\subFIT$. Fig. \ref{FIG:FilProf_pFIT_sgwO_sgSigB_pINT4} shows contours of constant $p\subFIT$ on the ($\sigma_{_{w\subO}},\sigma_{_{\Sigma\subB}}\!$) plane, for filaments with $p\subINT =4$; similarly Figs. \ref{FIG:FilProf_pFIT_sgwO_sgSigB_pINT3} and \ref{FIG:FilProf_pFIT_sgwO_sgSigB_pINT2} show the analogous results for $p\subINT =3$ and $p\subINT =2$, respectively. In each case we see that $w\subO$ is the primary parameter; at fixed $\sigma_{_{\Sigma\subB}}\!,$ $p\subFIT$ decreases monotonically and relatively rapidly with increasing $\sigma_{_{w\subO}}\!,$ especially for larger $p\subINT$. $\,\Sigma\subB$ is the secondary parameter: at fixed but finite $\sigma_{_{w\subO}}$, $p\subFIT$ decreases monotonically but relatively slowly with increasing $\sigma_{_{\Sigma\subB}}$, and tends towards a constant asymptotic value.

Finally, if we vary the range of $\Sigma\subO$ (i.e. vary $\sigma_{_{\Sigma\subO}}\!$) and simultaneously vary the range of $w\subO$ (i.e. vary $\sigma_{_{w\subO}}$), with $\Sigma\subB$ fixed at it default value (i.e. $\Sigma\subB =60\,\rm{M_{_\odot}\,pc^{-2}}$), we find that $p\subFIT$ is completely independent of the value of $\sigma_{_{\Sigma\subO}}\!$, and depends on $\sigma_{_{w\subO}}$ in exactly the same way as when $\sigma_{_{w\subO}}$ was varied on its own, i.e. as shown on Fig. \ref{FIG:FilProf_pFIT_sgwO}. Thus $\Sigma\subO$ is a null parameter: under no circumstance does its range have an effect on $p\subFIT$.

\subsection{Effect of correlation between $\Sigma\subO$ and $\Sigma\subB$}

If we include the correlation between $\Sigma\subO$ and $\Sigma\subB$ (i.e. we generate values of $\Sigma\subO$ using Eq. \ref{EQN:correlation01} rather than Eq. \ref{EQN:SigmaO.L}), the results are unchanged. This is unsurprising, since $\Sigma\subO$ is the null parameter.

\subsection{Comparison with observational data}

The numbers in square brackets on Figs. \ref{FIG:FilProf_pFIT_sgwO_sgSigB_pINT4}, \ref{FIG:FilProf_pFIT_sgwO_sgSigB_pINT3} and \ref{FIG:FilProf_pFIT_sgwO_sgSigB_pINT2} show the values of $\sigma_{_{w\subO}}$ and $\sigma_{_{\Sigma\subB}}$ estimated for the different regions analysed by \citet{ArzoumanianDetal2019}, as per the key in the caption to Fig. \ref{FIG:FilProf_pFIT_sgwO_sgSigB_pINT4} and Table A1 in Appendix \ref{APP:Arzoumanian}.

As treated here, the effect we have evaluated appears insufficient to reduce $p\subINT$ from $p\subINT =4.0$ \citep[as appropriate for an isolated, infinitely-long, isothermal filament in hydrostatic equilibrium,][]{OstrikerJ1964} to $p\subFIT \sim 2$ \citep[as reported by e.g.][]{ArzoumanianDetal2011, PalmeirimPetal2013, AndrePetal2014, FederrathCetal2016, ArzoumanianDetal2019}. However, we should note that \citet{HowardAetal2019} find -- on the basis of high-resolution maps derived using PPMAP -- that they can obtain a better fit to small local sections of the L1495 filament in Taurus with $p\subFIT =4$ rather than $p\subFIT =2$.

A more accurate evaluation should take into account two factors. First, the averaging applied by \citet{ArzoumanianDetal2019} only involves the profiles for individual filaments, and the range of $w\subO$ values for an individual filament is likely to be lower than the range for the ensemble of all the filaments in a particular region. Correcting for this will decrease $\Delta p$. Second, the averaging over the ensemble of filaments will reduce the range of $w\subO$ values, as shown by \citet{PanopoulouG2017}, and correcting for this will increase $\Delta p$. $\;\Delta p$ will always be positive, so $p\subFIT$ will always be less than $p\subINT$.

\section{Conclusions}\label{SEC:Conc}

We have shown that averaging filament profiles can reduce the fitted Plummer exponent, $p\subFIT$ below its intrinsic value, $p\subINT$, i.e. it artificially reduces the slope of the surface-density profile at large distance from the spine. (It is tempting to speculate that this effect operates even if the intrinsic surface-density profile is not well fit by a Plummer profile, but we have not proven this.)

The amount of reduction is largely determined by the intrinsic Plummer exponent, $p\subINT$, and the range of Plummer scale-lengths, $w\subO$, with a small additional contribution from the range of background surface-densities, $\Sigma\subB$.

This reduction is not affected by the apparent correlation between the the spinal surface-density, $\Sigma\subO$ and the background surface-density, $\Sigma\subB$, as reported in \citet{ArzoumanianDetal2019}. In Appendix A we explore the causes of this correlation, and suggest that it may be largely a selection effect.

For the ranges reported by \citet{ArzoumanianDetal2019}, the effect we have evaluated cannot, on its own, support values of $p\subINT$ significantly greater than $p\subFIT =2$. Specifically, it appears that $p\subINT\simeq 4.0$ is only reduced to $p\subFIT\gtrsim 3.0$ (see numbers in square brackets on Fig. \ref{FIG:FilProf_pFIT_sgwO_sgSigB_pINT4}), and $p\subINT\simeq 3.0$ is only reduced to $p\subFIT\gtrsim 2.5$ (numbers on Fig. \ref{FIG:FilProf_pFIT_sgwO_sgSigB_pINT3}).

However, if there were some other effect that operated in tandem with the one we have evaluated, then values of $p\subINT$ significantly greater than $p\subFIT\simeq 2$ might be plausible.

\section*{Data Availability}
The provenance of the data used in this paper is described in \citet{ArzoumanianDetal2019}. All software used will be supplied on request to APW.

\section*{Acknowledgements}
APW and FDP gratefully acknowledge the support of an STFC Consolidated Grant (ST/K00926/1). DA acknowledges financial support from the CNRS. We thank the referee for a careful and constructive report on the initial version of the paper which resulted in significant refinements.

\bibliographystyle{mnras}
\bibliography{FilProfFit}

\appendix

\section{The Arzoumanian et al. (2019) data-set}\label{APP:Arzoumanian}

\begin{figure}
\vspace{-1.20cm}\hspace{-5.10cm}\includegraphics[angle=270.0,width=2.35\columnwidth]{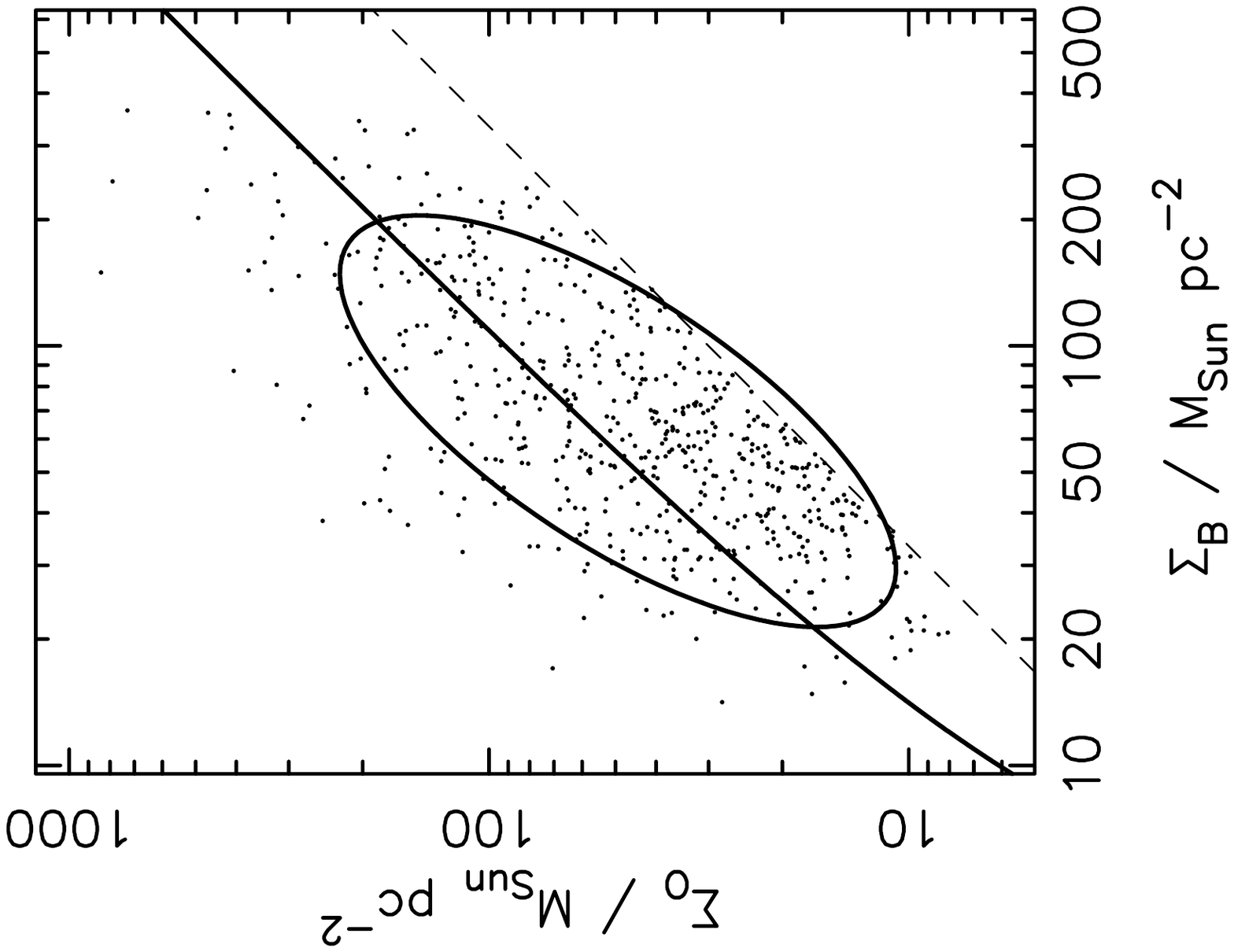}
\vspace{-0.0cm}
\caption{The distribution of profile parameters on the $(\sigma_{_{\Sigma\subB}},\sigma_{_{\Sigma\subO}})$ plane, and the corresponding moment ellipse (see text for definition).The solid straight line shows the correlation determined by \citet{ArzoumanianDetal2019}, and the dashed line marks the contrast threshold used by \citet{ArzoumanianDetal2019} to define a filament.}
\label{FIG:FilProfFit_SigmaB.SigmaO_AllProfiles}
\end{figure}

\begin{figure}
\vspace{-1.20cm}\hspace{-5.10cm}\includegraphics[angle=270.0,width=2.35\columnwidth]{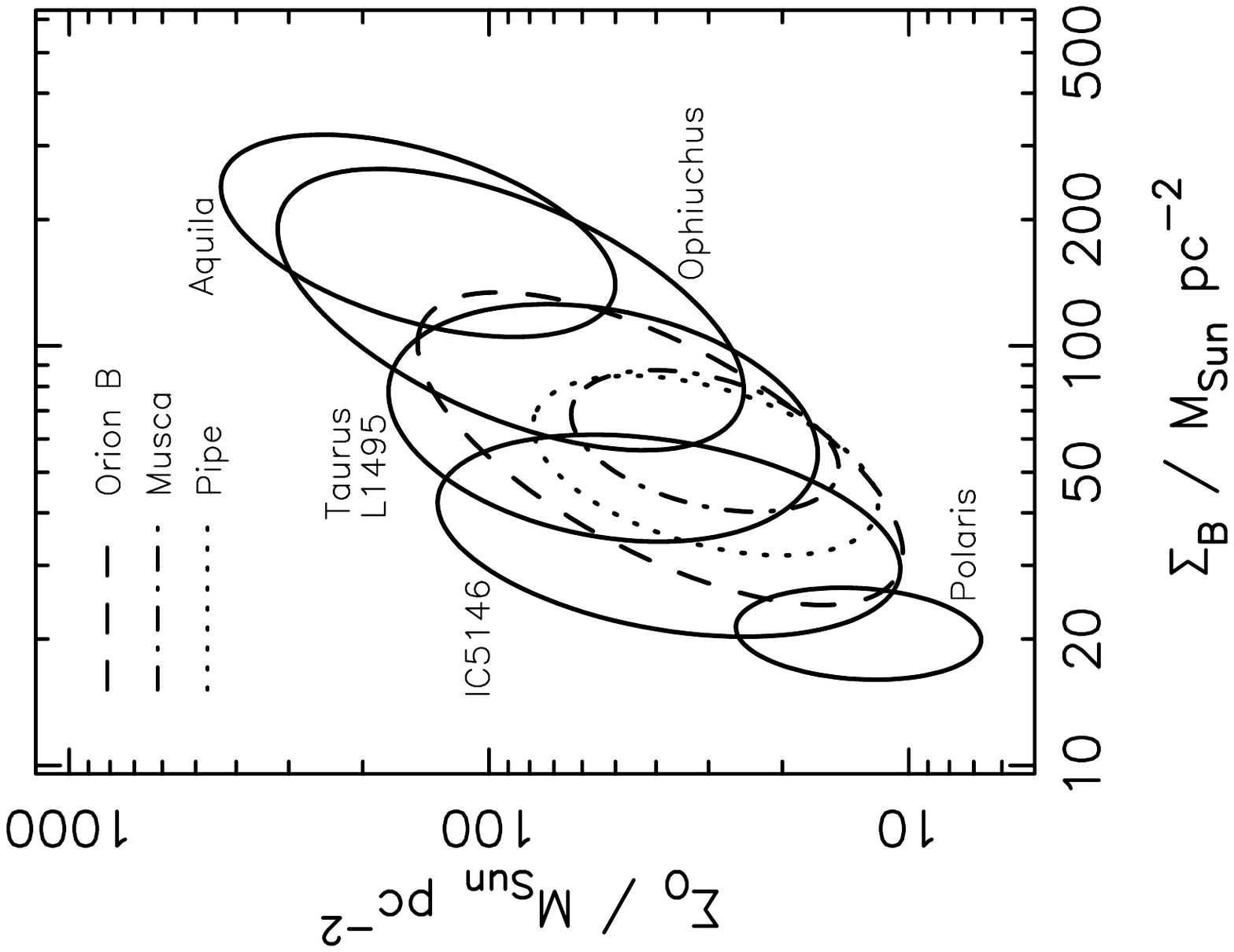}
\vspace{-0.0cm}
\caption{Moment ellipses on the $(\sigma_{_{\Sigma\subB}},\sigma_{_{\Sigma\subO}})$ plane for the different fields.}
\label{FIG:FilProfFit_SigmaB.SigmaO_AllEllipses}
\end{figure}

\begin{table*}
\begin{center}
\caption{Column 1 gives the name of the field considered, preceded by the number (in square brackets) used to represent this field on Figs.  \ref{FIG:FilProf_pFIT_sgwO_sgSigB_pINT4}, \ref{FIG:FilProf_pFIT_sgwO_sgSigB_pINT3} and \ref{FIG:FilProf_pFIT_sgwO_sgSigB_pINT2}. Column 2 gives the number of filaments. Columns 3 and 4 give the mean, $\mu_{_{\mbox{\sc fwhm}}}$, and standard deviation, $\sigma_{_{\mbox{\sc fwhm}}}$, of $\log_{_{10}}(\mbox{\sc fwhm}/\rm{pc})$, where $\mbox{\sc fwhm}$ is the full-width at half-maximum. Columns 5 and 6 give the mean, $\mu_{_{\Sigma\subB}}\!$, and standard deviation, $\sigma_{_{\Sigma\subB}}\!$, of $\log_{_{10}}(\Sigma\subB/\rm{M_{_\odot}pc^{-2}})$, where $\Sigma\subB$ is the background surface-density. Columns 7 and 8 give the mean, $\mu_{_{\Sigma\subO}}\!$, and standard deviation, $\sigma_{_{\Sigma\subO}}\!$, of $\log_{_{10}}(\Sigma\subO/\rm{M_{_\odot}pc^{-2}})$, where $\Sigma\subO$ is the spinal surface-density. Columns 9 through 11 refer to the moments of the distribution of profile parameters on the $(\sigma_{_{\Sigma\subB}}\!,\sigma_{_{\Sigma\subO}}\!)$ plane: column 9 gives the slope of the principal axis, columns 10 and 11 give, respectively, the major and minor axes of the moment ellipse.} 
\begin{tabular}{lrcccccccccccccc}\hline
{\sc Field} & ${\cal N}\,$ & $\;\;\;\;$ & $\!\!\mu_{_{\mbox{\sc fwhm}}}\!\!$ & $\!\!\sigma_{_{\mbox{\sc fwhm}}}\!\!$ & $\;\;\;\;$ & $\mu_{_{\Sigma\subB}}$ & $\sigma_{_{\Sigma\subB}}$ & $\;\;\;\;$ & $\mu_{_{\Sigma\subO}}$ & $\sigma_{_{\Sigma\subO}}$ & $\;\;\;\;$ & $S$ & $\;\;$ & $a$ & $b$ \\\hline
{\sc [0] All} & 599 & & -0.96 & 0.19 & & 1.77 & 0.28 & & 1.64 & 0.29 & & 0.198 & & 0.770 & 0.295 \\\hline
{\sc [1] IC5146} & 59 & & -0.80 & 0.17 & & 1.49 & 0.14 & & 1.52 & 0.32 & & 5.82 & & 0.558 & 0.225 \\
{\sc [2] Orion B} & 234 & & -0.82 & 0.12 & & 1.70 & 0.22 & & 1.54 & 0.33 & & 1.88 & & 0.641 & 0.250 \\
{\sc [3] Aquila} & 71 & & -1.05 & 0.15 & & 2.21 & 0.14 & & 2.11 & 0.27 & & 3.27 & & 0.488 & 0.203 \\
{\sc [4] Musca} & 10 & & -1.10 & 0.24 & & 1.72 & 0.10 & & 1.43 & 0.18 & & 3.86 & & 0.327 & 0.153 \\
{\sc [5] Polaris} & 20 & & -1.15 & 0.13 & & 1.26 & 0.06 & & 1.06 & 0.17 & & 16.7 & & 0.293 & 0.108 \\
{\sc [6] Pipe} & 38 & & -1.10 & 0.17 & & 1.66 & 0.12 & & 1.43 & 0.24 & & 3.49 & & 0.519 & 0.269 \\
{\sc [7] Taurus L1495} & 110 & & -1.15 & 0.20 & & 1.76 & 0.16 & & 1.67 & 0.30 & & 5.02 & & 0.519 & 0.269 \\
{\sc [8] Ophiuchus} & 57 & & -1.15 & 0.13 & & 2.03 & 0.19 & & 1.89 & 0.32 & & 2.28 & & 0.600 & 0.257 \\\hline
\end{tabular}
\end{center}
\label{TAB:Means}
\end{table*}

Table A1 gives values of the distribution parameters derived from the large sample analysed by \citet{ArzoumanianDetal2019}, viz. ${\cal N}$ (the number of filaments analysed); $\mu_{_{\mbox{\sc fwhm}}}$ and $\sigma_{_{\mbox{\sc fwhm}}}$ (the mean and standard deviation for the logarithm of the full-width at half-maximum); $\mu_{_{\Sigma\subB}}$ and $\sigma_{_{\Sigma\subB}}$ (the mean and standard deviation for the background surface-density); and $\mu_{_{\Sigma\subO}}$ and $\sigma_{_{\Sigma\subO}}$ (the mean and standard deviation for the spinal surface-density). The values of $\sigma$ are taken at their face value, although we note the arguments in \citet{PanopoulouG2017} suggesting that they may be underestimates. (Since for a given $p\subINT$, the {\sc fwhm} is proportional to $w\subO$, we take the standard deviation of the logarithm of $w\subO$ to be the same as the the standard deviation of the logarithm of {\sc fwhm}.)

Fig. \ref{FIG:FilProfFit_SigmaB.SigmaO_AllProfiles} shows the values of $\Sigma\subO$ and $\Sigma\subB$ for individual filaments. This is essentially the same as Fig. 6c in \citet{ArzoumanianDetal2019}, except that the axes are $\Sigma$ rather than $N_{_{\rm H_2}}$ (see Eq. \ref{EQN:NH2.2.Sigma.01}), and they are scaled equally. We have over-plotted (a) with a solid line the correlation derived by \citet{ArzoumanianDetal2019},
\begin{eqnarray}
\Sigma\subO&=&0.95\,\Sigma\subB\,-\,3.4\,\rm{M_{_\odot}\,pc^{-2}},
\end{eqnarray}
(b) with a dashed straight line the contrast threshold adopted by \citet{ArzoumanianDetal2019},
\begin{eqnarray}\label{EQN:CO.01}
C\subO\;\,\equiv\;\,\Sigma\subO/\Sigma\subB&>&0.3,
\end{eqnarray}
and (c) with a solid line the `moment ellipse'. 

The moment ellipse is the ellipse which, if the same number of points were distributed uniformly within its boundary, would have the same centre of mass as the actual points, the same principal moments and the same principal axes. This is an alternative way of displaying a linear correlation between two variables (or their logarithms). It has the merit that it treats the two variables equivalently, i.e. it does not assume that one is dependent and the other independent.

There are two key things to note about this plot. First, the column-density contrast threshold that \citet{ArzoumanianDetal2019} apply,
accounts for the lower cut-off in values of $\Sigma\subO$, and is probably responsible for a significant part of the observed correlation. Second the empty bit of the ellipse below this threshold is to some extent compensated by a concentration of points immediately above the threshold.

Fig. \ref{FIG:FilProfFit_SigmaB.SigmaO_AllEllipses} shows the moment ellipses for the individual fields. It shows that within individual fields, the correlation between $\Sigma\subO$ and $\Sigma\subB$ varies. First, it is always steeper than for the ensemble of all the fields. In general,  the spread of $\Sigma\subO$ values is larger than the spread of $\Sigma\subB$ values. Second, the cases where this is less marked are those more affected by the contrast threshold. Third, the extent of the correlation noted by \citet{ArzoumanianDetal2019} can be attributed to systematic changes from one field to another, with Polaris at one extreme,  Aquila at the other extreme, and Musca and Pipe in the middle (and having very little overlap with either Polaris or Aquila).

\bsp	
\label{lastpage}
\end{document}